\documentclass[twocolumn]{article}
\usepackage{spconfa4,amsmath,graphicx}
\usepackage{bm,amssymb,siunitx,booktabs,cleveref,url,cite}
\usepackage{subcaption}

\title{Neural Audio Codec with Adjustable Token Temporal Resolution Using Sampling-Frequency-Independent Convolutional Layers}
\name{Tomohiko Nakamura${}^\dagger$\thanks{This work was supported by JSPS KAKENHI under Grant JP23K28108 and JST PRESTO under Grant number JPMJPR2517.}, Wataru Nakata${}^\ddagger$, Kanami Imamura${}^{\dagger,\ddagger}$, Yuki Saito${}^\ddagger$}
\address{
    ${}^\dagger$ The National Institute of Advanced Industrial Science and Technology (AIST), \\
    2-4-7 Aomi, Koto-ku, Tokyo, 135-0064 Japan \\
    ${}^\ddagger$ The University of Tokyo, 7-3-1 Hongo, Bunkyo-ku, Tokyo, 113-8654 Japan
}

\crefname{figure}{Fig.}{Figs.}
\Crefname{figure}{Figure}{Figures}

\newcommand{\LAFparam}{\theta_{c,c'}}
\newcommand{\ltrin}{\Delta_{\text{tok}}^{\text{in}}}
\newcommand{\ltrout}{\Delta_{\text{tok}}^{\text{out}}}
\newcommand{\vltrin}{\tilde{\Delta}_{\text{tok}}^{\text{in}}(S)}
\newcommand{\vltrout}{\tilde{\Delta}_{\text{tok}}^{\text{out}}}
\newcommand{\bks}{K_{\mathrm{base}}}

\allowdisplaybreaks

\begin{document}
\ninept
\maketitle
\begin{abstract}
Discrete tokens obtained from neural audio codecs (NACs) have been used as compact representations in audio generation and understanding models. In such token-based systems, token temporal resolution (TTR), defined as the time interval between adjacent token frames, is important because it controls the trade-off between representing rapid acoustic events and reducing token-sequence length. However, most NACs are trained at a single TTR and require separate training for each TTR. This paper proposes a mechanism that enables a single NAC to operate at multiple TTRs using sampling-frequency-independent convolutional layers. The mechanism regards TTR as the sampling period of the token sequence and generates TTR-dependent convolutional kernels from a shared parameter set, while adjusting the kernel size and stride for each TTR. We incorporate the mechanism into Descript Audio Codec, leaving the quantizer unchanged. Experiments on environmental sound reconstruction show that the proposed model outperforms a single-model baseline that switches TTR-specific layers for each TTR.
\end{abstract}
\section{Introduction}
\label{sec:intro}
Neural audio codecs (NACs) encode audio signals into discrete tokens and reconstruct waveforms from them~\cite{Mousavi2025TMLR}.
While conventional codecs such as Opus~\cite{Opus} are primarily designed for audio compression, NACs have gathered attention because their learned tokens can also serve as compact representations for other audio models.
For example, NAC tokens have been used as inputs to audio generation models~\cite{Borsos2023IEEEACMTASLP,Agostinelli2023arXiv} and audio understanding models~\cite{Rubenstein2023arXiv,Zhang2024ICLR}.
This broader use of NAC tokens makes the design of token sequences important.

Among the design factors of token sequences, token temporal resolution (TTR), defined as the time interval between adjacent token frames, is crucial.
A finer TTR can capture short acoustic events and rapid temporal changes, whereas a coarser TTR shortens the token sequence and is better suited for long-context modeling.
This trade-off is relevant to both reconstruction quality and the computational cost of downstream token-based modeling.
Recent low-frame-rate tokenizers and codecs have highlighted the importance of this trade-off across speech, music, and general audio domains~\cite{Ji2025ICLR,Liu2024JSTSP,Li2025Interspeech,Casanova2025Interspeech,Li2026ICLR}.
Since the appropriate TTR depends on the target application and the temporal characteristics of the audio, a single codec should support multiple TTRs without requiring a separate model for each resolution.

Supporting multiple TTRs in a single codec amounts to sharing model parameters across token sequences with different temporal resolutions.
We focus on the fact that the TTR of a token sequence can be interpreted as its sampling period.
This interpretation connects TTR control to sampling-frequency-independent (SFI) convolutional layers, which have been developed in the context of audio source separation~\cite{KSaito2022IEEEACMTASLP,KImamura2024APSIPATrans,KImamura202511ESP}.
These layers handle signals with different sampling periods by generating convolutional weights according to the input sampling period.
To this end, their parameters are defined in a domain independent of the sampling period rather than at a fixed discrete-time resolution.

In this paper, we propose a mechanism that enables a single NAC to support multiple TTRs using this parameterization.
In a typical NAC, the encoder produces a latent sequence that is fed to the quantizer, and the temporal resolution of this sequence determines the TTR of the resulting tokens.
The proposed mechanism uses SFI convolutional layers to change the temporal resolution of the latent sequence.
Unlike conventional applications of SFI convolutional layers, where weights are generated according to the input sampling period, the proposed mechanism conditions weight generation on a token-level sampling period determined from the target TTR.
For each target TTR, the mechanism adjusts the kernel size and stride of these layers and obtains their weights using the SFI parameterization.
We incorporate the mechanism in Descript Audio Codec (DAC)~\cite{Kumar2023NIPS}, leaving the quantizer unchanged.
Through sound reconstruction experiments, we evaluate its reconstruction quality across various TTRs.
To the best of our knowledge, this is the first attempt to introduce SFI convolutional layers into a NAC.

\section{Related Work} \label{sec:related}
\subsection{NAC} \label{sec:nac}
A typical NAC consists of an encoder, a quantizer, and a decoder.
The encoder maps an input waveform into a latent sequence, the quantizer converts this sequence into discrete tokens, and the decoder reconstructs the waveform from the tokens.
SoundStream established this framework using convolutional encoder--decoder networks and residual vector quantization (RVQ)~\cite{Zeghidour2021IEEEACMTASLP_SoundStream}.
It trains the entire model using reconstruction and adversarial losses.
RVQ quantizes the latent sequence through multiple vector-quantization (VQ) stages, each of which encodes the residual error left by the previous stage.
EnCodec improves neural audio compression with a streaming architecture, a multiscale spectrogram adversary, and a loss-balancing strategy~\cite{Defossez2023TMLR}.
DAC follows the same architecture while improving the RVQ design and adversarial training~\cite{Kumar2023NIPS}.
DAC is widely used as a baseline for audio tokenization~\cite{Mousavi2025TMLR}, and we adopt it as the base architecture in this paper.

A few NACs have been designed to handle multiple TTRs within a single model.
Multiscale RVQ incorporates multiple TTRs inside the quantizer by using VQs at different temporal resolutions~\cite{Siuzdak2024NIPSW}, while low-frame-rate and variable-frame-rate codecs use rate-dependent mechanisms to operate at multiple or variable TTRs~\cite{Zhang2025Interspeech,Li2026ICLR}.
These studies show the usefulness of moving beyond fixed-TTR codecs.
However, their support for multiple or variable TTRs relies on rate-specific processing.
In contrast, the proposed mechanism treats TTR as an input condition to a shared parameterization.
This design provides a unified way to handle multiple TTRs without preparing separate parameter sets for each TTR.

\subsection{DAC} \label{sec:dac}
DAC follows the encoder--quantizer--decoder architecture and uses convolutional layers in the encoder and decoder~\cite{Kumar2023NIPS}.
The encoder first expands the channel dimension of the input waveform and then applies a stack of downsampling blocks, each of which contains residual subnetworks with Snake activations~\cite{Liu2020NeurIPS} followed by a strided convolutional layer.
At each VQ stage, the RVQ-based quantizer projects the current residual to a lower-dimensional space and performs VQ using cosine distance.
The decoder has an upsampling structure symmetric to the encoder, where transposed convolutional layers increase the temporal resolution.
The final output is passed through a hyperbolic tangent function to constrain the reconstructed waveform to the range $[-1,1]$.
Weight normalization~\cite{Salimans2016NIPS} is applied to the convolutional layers in the encoder and decoder.

DAC is trained using a combination of a reconstruction loss, VQ-related losses, adversarial losses, and a feature-matching loss~\cite{Kumar2023NIPS}.
The reconstruction loss is a multi-scale mel-spectrogram loss computed from log-mel spectrograms at multiple time--frequency resolutions.
The VQ-related losses consist of the codebook and commitment losses introduced for vector-quantized variational autoencoders~\cite{vanDenOord2017NIPS}.
For adversarial training, DAC uses a multi-period discriminator in the waveform domain and a multi-band multi-scale short-time Fourier transform (STFT) discriminator in the spectral domain, together with a feature-matching loss.

\section{Proposed Method} \label{sec:proposed}
This section describes the proposed TTR-adjustment mechanism.
The mechanism makes the TTR adjustable by using SFI convolutional layers and by setting their sampling period, kernel size, and stride according to the target TTR.
We first review SFI convolutional layers and then describe the incorporation of this mechanism into DAC as a base architecture.

\subsection{SFI Convolutional Layers} \label{sec:sfi_conv}
The SFI convolutional layer is a one-dimensional (1D) temporal convolutional layer that extends an ordinary convolutional layer by replacing its fixed weights with weights generated for a specified sampling period~\cite{KSaito2022IEEEACMTASLP}.
A 1D convolutional layer computes cross-correlation between input features and its weights.
When the weights are time-reversed, this cross-correlation coincides with the convolution operation in signal processing.
Thus, a convolutional layer can be interpreted as a collection of digital filters whose discrete-time impulse responses are given by the time-reversed weights.

The key idea of the SFI convolutional layer is not to learn these discrete-time impulse responses directly.
Instead, the weight-generation module defines trainable functions in an SFI domain, i.e., a domain independent of the sampling period, and converts them into discrete-time impulse responses for the specified sampling period.
This weight-generation process is analogous to analog-to-digital filter conversion, and we refer to these trainable functions as latent analog filters.
This analogy allows us to use digital filter design methods for weight generation.

\begin{figure}[t]
  \centering
  \includegraphics[width=0.9\columnwidth,clip]{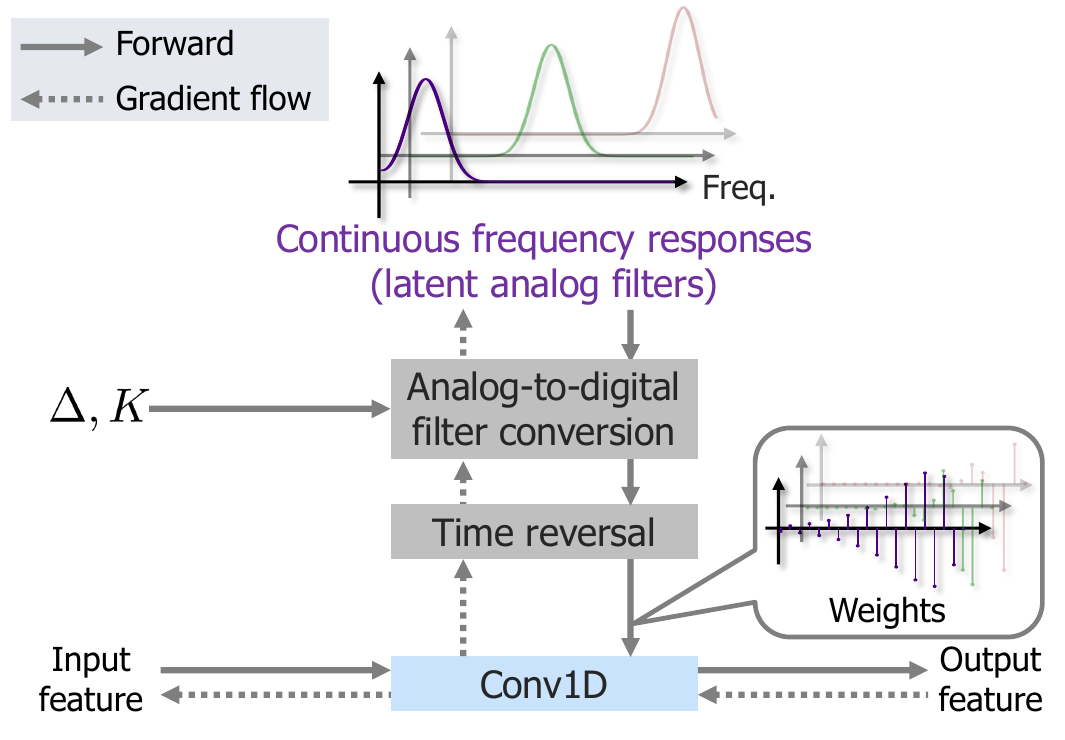}
  \caption{
    Overview of the SFI convolutional layer based on frequency-domain filter design.
    \textsf{Conv1D} denotes a 1D convolutional layer.
  }
  \label{fig:sfi_conv}
\end{figure}

\Cref{fig:sfi_conv} shows an overview of the SFI convolutional layer based on frequency-domain filter design.
In this method, each latent analog filter is represented as a continuous frequency response.
Let $C$ and $C'$ denote the numbers of input and output channels, respectively.
For an input--output channel pair $(c,c')$, let this response be $G_{c,c'}(\omega;\theta_{c,c'})$, where $\omega$ is the angular frequency and $\theta_{c,c'}$ denotes its trainable parameters.
For a given sampling period $\Delta$ and kernel size $K$, the weight-generation module evaluates $G_{c,c'}(\omega;\theta_{c,c'})$ at discrete frequency points up to the Nyquist frequency $\pi/\Delta$.
It then obtains a length-$K$ discrete-time impulse response by fitting its frequency response to the sampled values in a least-squares sense.
The impulse response is time-reversed and used as the weight of a 1D convolutional layer with stride $S$.
See~\cite{KSaito2022IEEEACMTASLP} for further details.

An SFI counterpart of a transposed convolutional layer (SFI transposed convolutional layer) can be defined in the same manner.
The same weight-generation module is used, while the generated weights are applied to a transposed convolutional layer instead of an ordinary convolutional layer.

\subsection{Incorporation of SFI Layers into DAC} \label{sec:dac_sfi_conv}
\begin{figure}[t]
  \centering
  \includegraphics[width=\columnwidth,clip]{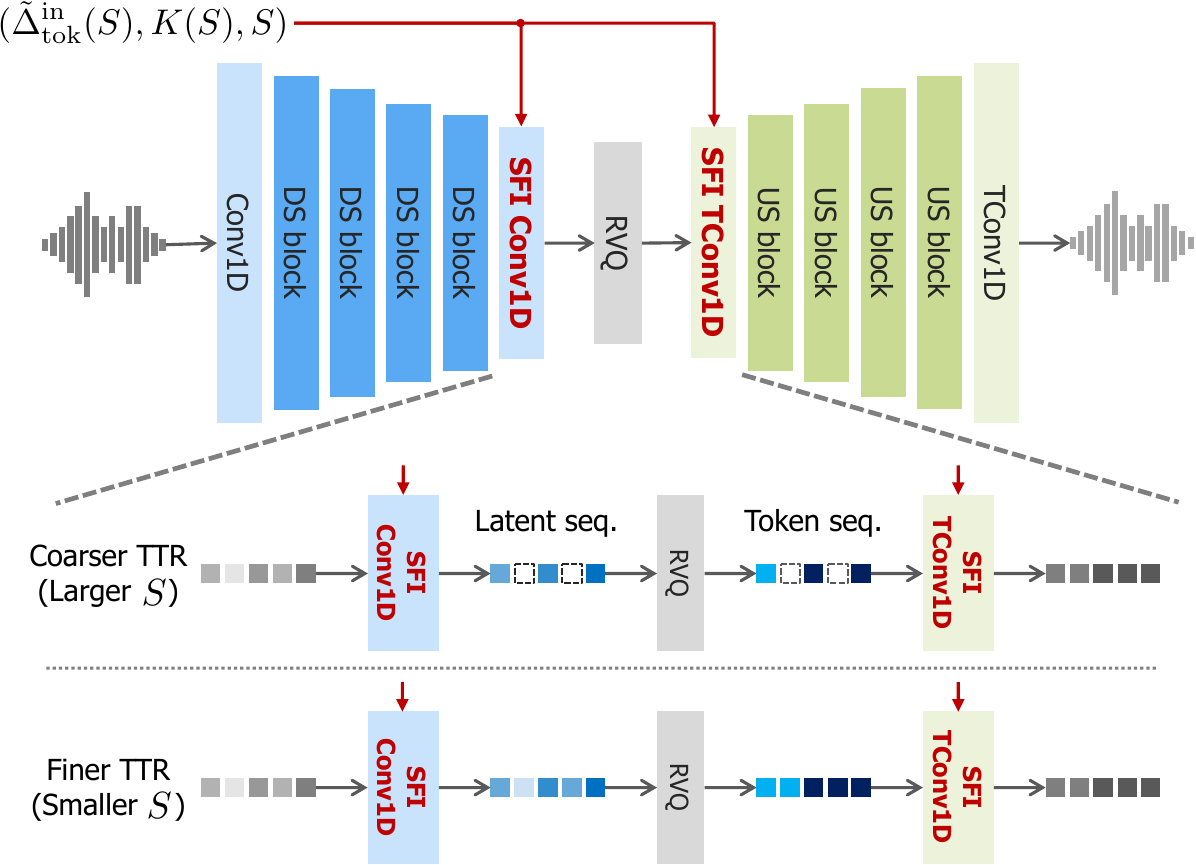}
  \caption{
    DAC architecture equipped with the proposed TTR-adjustment mechanism.
    The upper part shows the entire architecture, and the lower part illustrates two examples for different target TTRs.
    The notation \textsf{Conv1D} follows \cref{fig:sfi_conv}, and \textsf{TConv1D} denotes a 1D transposed convolutional layer.
    \textsf{SFI Conv1D} and \textsf{SFI TConv1D} denote their SFI counterparts.
    DS and US blocks denote downsampling and upsampling blocks, respectively.
    }
  \label{fig:proposed_dac}
\end{figure}

We incorporate the SFI layers into DAC so that the weights of the layers around the quantizer can be generated according to the target TTR.
In DAC, since the quantizer does not change the temporal resolution, the TTR is determined by the temporal resolution of the encoder output.
This resolution can be interpreted as the sampling period of the encoder-output sequence.
With this interpretation, we change only the layers immediately before and after the quantizer.

\Cref{fig:proposed_dac} shows the resulting DAC architecture.
We replace the last convolutional layer of the encoder with an SFI convolutional layer and the first transposed convolutional layer of the decoder with an SFI transposed convolutional layer.
SFI layers can replace the corresponding layers without changing the surrounding architecture, except that the sampling period is additionally provided for weight generation.
This allows the layers around the quantizer to generate weights according to the target TTR, while leaving the quantizer and the other encoder and decoder layers unchanged.
This preserves the original DAC quantization scheme and allows the model to be trained using the same training framework as DAC.

\subsection{Adjustment Method for Different TTRs} \label{sec:proposed_tr_conversion}
We now describe how to adjust the sampling period, stride, and kernel size of the SFI layers for a target TTR.
When SFI layers are applied in audio source separation, their kernel sizes and strides are adjusted according to the input sampling period~\cite{KSaito2022IEEEACMTASLP}.
This adjustment keeps the continuous-time durations corresponding to the kernel size and stride unchanged across different input sampling periods.
The sampling period provided to the weight-generation module is also set according to the sampling period of the waveform input to the separation network.
With these settings, the layer covers the same time range and produces outputs at the same temporal interval even when the waveform sampling period changes.
By contrast, in the proposed TTR-adjustment setting, the input sampling period $\ltrin$ of the SFI layer is fixed, while its output sampling period $\ltrout$ is changed according to the target TTR.
This difference prevents us from directly using the adjustment method developed for audio source separation.

To solve this problem, the proposed adjustment method keeps the token-level time scale for weight generation common across target TTRs.
We introduce a virtual token sampling period $\vltrout$, which is independent of the actual TTR $\ltrout$.
The output of the SFI convolutional layer is interpreted as a virtual token sequence sampled with period $\vltrout$.
When the SFI convolutional layer has stride $S$, its input is interpreted as a virtual sequence sampled with period $\vltrin$:
\begin{equation}
  \vltrin = \vltrout/S.
\end{equation}
This virtual input sampling period $\vltrin$ is used as the sampling period provided to the weight-generation module of the SFI layers.

The kernel size is also adjusted on this virtual time scale.
Let $\bks$ be the kernel size when $S=1$.
For stride $S$, the kernel size is given by
\begin{equation}
  K(S) = \bks S .
\end{equation}
This keeps $K(S)/S$ constant, meaning that the kernel covers the same number of virtual token intervals across different TTRs.

The actual TTR is realized by choosing the stride $S$ according to the actual input sampling period $\ltrin$.
For a target TTR $\ltrout$, $S$ is chosen to satisfy
\begin{equation}
  \ltrout = S\ltrin .
\end{equation}
Once $S$ is chosen, the virtual input sampling period $\vltrin$ and the kernel size $K(S)$ are determined as described above.

In summary, the SFI layers use $\vltrin$ as the sampling period $\Delta$ for weight generation, $K(S)$ as the kernel size, and $S$ as the stride (see \cref{fig:proposed_dac}).
For the SFI convolutional layer, the input and output correspond to virtual sequences with sampling periods $\vltrin$ and $\vltrout$, respectively.
For the SFI transposed convolutional layer, this relation is reversed: its input corresponds to the virtual token sequence with sampling period $\vltrout$, and its output corresponds to the virtual sequence with sampling period $\vltrin$.

One advantage of the proposed mechanism is that it modifies only the layers around the quantizer and leaves the other parts unchanged.
Although we have thus far derived the mechanism using DAC, the same idea can be applied to various NAC architectures.
For convolutional NACs, the mechanism can be introduced by replacing the convolutional layers immediately before and after the quantizer with SFI layers.
For NACs without such convolutional layers around the quantizer, the mechanism can instead be introduced by inserting SFI layers before and after the quantizer.
The mechanism does not assume a specific quantizer structure, as long as the temporal resolution of the latent sequence can be controlled around the quantizer.
In this paper, we focus on demonstrating the proposed mechanism using DAC as a representative case, so we leave experimental validation on other NAC architectures for future work.

\section{Experiments} \label{sec:exp}
\begin{table}[t]
  \centering
  \caption{Kernel sizes and strides of the layers adjacent to the quantizer for different TTRs}
  {
    \footnotesize
  \begin{tabular}{
    S[table-format=3.1]
    S[table-format=2.1]
    |
    S[table-format=2.0]
    S[table-format=2.0]
  } \toprule
    {TTR [\si{\milli\second}]} & {Token frame rate [\si{\hertz}]} & {$K$ [sample]} & {$S$ [sample]} \\
    \midrule
    13.3  & 75.0 & 7  & 1  \\
    26.7  & 37.5 & 14 & 2  \\
    40.0  & 25.0 & 21 & 3  \\
    53.3  & 18.8 & 28 & 4  \\
    66.7  & 15.0 & 35 & 5  \\
    106.7 & 9.4  & 56 & 8  \\
    133.3 & 7.5  & 70 & 10 \\
    \bottomrule
  \end{tabular}
  }
  \label{tab:conv_params}
\end{table}

\begin{figure*}[t]
  \centering
  \includegraphics[width=0.73\columnwidth]{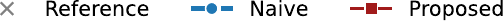}
  \\
  \begin{subfigure}{0.66\columnwidth}
    \centering
    \includegraphics[clip,width=\columnwidth]{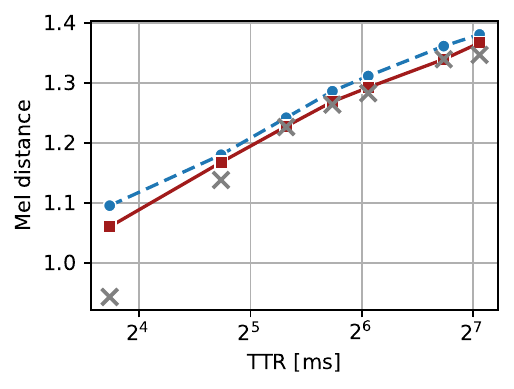}
    \caption{Mel distance}
  \end{subfigure}
  \hfill
  \begin{subfigure}{0.68\columnwidth}
    \centering
    \includegraphics[clip,width=\columnwidth]{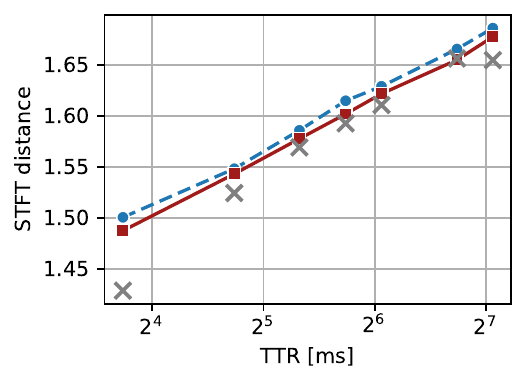}
    \caption{STFT distance}
  \end{subfigure}
  \hfill
  \begin{subfigure}{0.66\columnwidth}
    \centering
    \includegraphics[clip,width=\columnwidth]{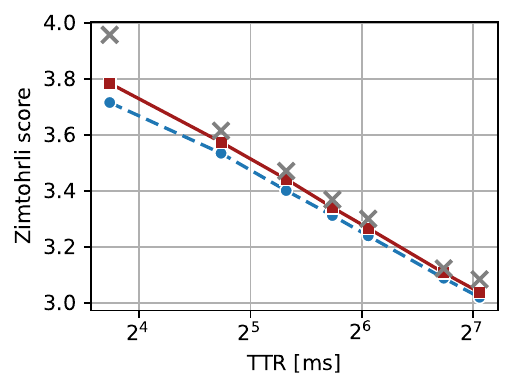}
    \caption{Zimtohrli score}
  \end{subfigure}
  \vspace{-2mm}
  \caption{Reconstruction performance at different TTRs for Reference, Naive, and Proposed. Lower mel and STFT distances indicate better reconstruction performance, and a higher Zimtohrli score indicates better perceptual quality.}
  \label{fig:results}
  \vspace{-1mm}
\end{figure*}

\subsection{Experimental Setup}
To evaluate the proposed mechanism, we conducted sound reconstruction experiments on the CochlScene dataset~\cite{Jeong2022APSIPA}.
This dataset is a crowdsourced environmental sound dataset consisting of monaural audio signals recorded in 13 acoustic scenes.
Each signal is \SI{10}{\second} long and originally sampled at \SI{44.1}{\kilo\hertz}.
The official split contains \num{60855} training samples (\SI{169.0}{\hour}), \num{7573} validation samples (\SI{21.0}{\hour}), and \num{7687} test samples (\SI{21.35}{\hour}).
We used this split without modification, except that all signals were resampled to \SI{24}{\kilo\hertz} to reduce computational cost.

We compared two single-model methods that handle the seven TTRs listed in \Cref{tab:conv_params}: a single-model baseline using TTR-specific layers (\textit{Naive}) and the proposed model using the SFI layers (\textit{Proposed}).
In addition, we trained TTR-specific DAC models as reference models (\textit{Reference}).
For all models, the kernel sizes and strides of the layers adjacent to the quantizer were set according to \Cref{tab:conv_params}.
Specifically, these layers correspond to the last convolutional layer in the encoder and the first transposed convolutional layer in the decoder for Naive and Reference, and to the SFI layers for Proposed.
Except for the model-specific layers described below, all hyperparameters were taken from the official DAC configuration for a sampling frequency of \SI{24}{\kilo\hertz}\footnote{\url{https://github.com/descriptinc/descript-audio-codec}}.

\smallskip
\noindent \textbf{Naive}. Naive is a single-model baseline that handles multiple TTRs by switching TTR-specific layers instead of using SFI layers.
Specifically, for each TTR, it has a separate last convolutional layer in the encoder and a separate first transposed convolutional layer in the decoder.
Once a TTR is chosen, the corresponding pair is used.

\noindent \textbf{Proposed}. Proposed is a DAC-based model equipped with the proposed mechanism.
It uses SFI layers adjacent to the quantizer to adjust the TTR.
The virtual token sampling period was set to $\vltrout=1/75$.
As the latent analog filters of the SFI layers, we used modulated Gaussian functions~\cite{KSaito2022IEEEACMTASLP}:
\begin{equation}
  G_{c,c'}(\omega;\LAFparam)
  =
  \sum_{l\in\{-1,1\}}
  \exp\left[
    -\frac{(\omega-l\mu_{c,c'})^2}{2p_{c,c'}^2}
    + jl\varphi_{c,c'}
  \right],
  \label{eq:mgf}
\end{equation}
where $\LAFparam=\{\mu_{c,c'},p_{c,c'},\varphi_{c,c'}\}$.
Here, $\mu_{c,c'}$ is the center angular frequency, $p_{c,c'}$ controls the bandwidth of the Gaussian component, and $\varphi_{c,c'}$ is the initial phase.
The center angular frequencies were initialized up to the Nyquist frequency of the virtual token frame rate: $\mu_{c,c'} = 75\pi(c'-1)/(C'-1)$.
The initial value of $p_{c,c'}$ was set to $1.5\pi$, and $\varphi_{c,c'}$ was initialized uniformly over $[0,\pi]$.
Naive and Proposed differed only in whether the adjacent layers were implemented as TTR-specific layers or SFI layers.

\noindent \textbf{Reference}. Reference DAC models were trained separately for each TTR to provide TTR-specific reference performance.

\smallskip

All models were trained using the training configuration of the official DAC implementation with the following modifications.
For Naive and Proposed, one of the TTRs was uniformly sampled for each batch throughout training.
The audio segment length in each batch was set to \SI{8}{\second} because we empirically found that longer segments improved reconstruction performance.
The weight of the multi-scale mel-spectrogram loss was changed from 15 to 30 to stabilize training.

Following \cite{Kumar2023NIPS}, we used the multi-scale mel-spectrogram loss (mel distance) and the multi-scale STFT loss (STFT distance) between the input and reconstructed signals as objective reconstruction metrics.
To complement these spectral metrics, we additionally used Zimtohrli, a perceptually motivated full-reference audio similarity metric~\cite{Alakuijala2025arXiv}.
The Zimtohrli score ranges from 1 to 5 and was computed using the official implementation\footnote{\url{https://github.com/google/zimtohrli}}.

\subsection{Results} \label{sec:results}
\Cref{fig:results} shows the average mel distance, STFT distance, and Zimtohrli score at each TTR.
Proposed consistently outperformed Naive across all TTRs in all three metrics while using fewer trainable parameters for the layers adjacent to the quantizer.
For each input--output channel pair, the number of trainable parameters in the SFI layers of Proposed is 3, whereas that in the TTR-specific layers of Naive is the sum of the kernel sizes over all TTRs, i.e., 231.
These results demonstrate that generating TTR-dependent kernels from a common TTR-independent representation is more effective and parameter-efficient than learning independent TTR-specific convolutional kernels.

Compared with Reference, Proposed achieved similar performance at larger TTRs, even though it handles multiple TTRs with a single model.
This indicates that the proposed mechanism can adjust the TTR without training a separate NAC for each TTR when the target TTR is relatively large.
As the TTR became smaller, however, the gap between Proposed and Reference increased.
This may be because the use of codebooks shared across TTRs limits the quantizer’s ability to represent finer token sequences.
Further investigation of TTR-dependent quantization is left for future work.

\section{Conclusion}
In this paper, we proposed a mechanism that enables a single NAC to operate at multiple TTRs using SFI layers.
The mechanism generates TTR-dependent weights from shared trainable parameters and modifies only the layers adjacent to the quantizer.
Experiments on an environmental sound dataset showed that the proposed model outperformed a single-model baseline with TTR-specific layers, despite having fewer trainable parameters.
The proposed model also achieved performance comparable to TTR-specific DAC reference models at larger TTRs.
Improving performance at smaller TTRs through TTR-dependent quantization remains future work.

\newpage

\bibliographystyle{IEEEbib}
\bibliography{abbrev,refs}

\end{document}